\documentclass[twocolumn,showpacs,preprintnumbers,amsmath,amssymb,prl]{revtex4}
\usepackage{graphicx}% Include figure files
\usepackage{bm}% bold math
\usepackage[usenames]{color}

\begin{document}
\title{Dynamic transition in supercritical iron}
\author{Yu. D. Fomin}
\affiliation{Institute for High Pressure Physics, Russian Academy
of Sciences, Troitsk 142190, Moscow, Russia \\ Moscow Institute of
Physics and Technology, Dolgoprudny, Moscow Region 141700, Russia}
\author{V. N. Ryzhov}
\affiliation{Institute for High Pressure Physics, Russian Academy
of Sciences, Troitsk 142190, Moscow, Russia \\ Moscow Institute of
Physics and Technology, Dolgoprudny, Moscow Region 141700, Russia}
\author{E. N. Tsiok}
\affiliation{Institute for High Pressure Physics, Russian Academy of Sciences, Troitsk 142190, Moscow Region, Russia}
\author{V. V. Brazhkin}
\affiliation{Institute for High Pressure Physics, Russian Academy
of Sciences, Troitsk 142190, Moscow Region, Russia}
\author{K. Trachenko}
\affiliation{School of Physics and Astronomy Queen Mary University of London, Mile End Road, London E1 4NS, United Kingdom}

\begin{abstract}
Recent advance in understanding the supercritical state posits the
existence of a new line above the critical point separating two
physically distinct states of matter: rigid liquid and non-rigid
gas-like fluid. The location of this line, the Frenkel line,
remains unknown for important real systems. Here, we map the
Frenkel line on the phase diagram of supercritical iron using
molecular dynamics simulations. On the basis of our data, we
propose a general recipe to locate the Frenkel line for any
system, the recipe that importantly does not involve
system-specific detailed calculations and relies on the knowledge
of the melting line only. We further discuss the relationship
between the Frenkel line and the metal-insulator transition in
supercritical liquid metals. Our results enable predicting the
state of supercritical iron in several conditions of interest. In
particular, we predict that liquid iron in the Jupiter core is in
the ``rigid liquid'' state and is highly conducting. We finally
analyse the evolution of iron conductivity in the core of smaller
planets such as Earth and Venus as well as exoplanets: as planets
cool off, the supercritical core undergoes the transition to the
rigid-liquid conducting state at the Frenkel line.
\end{abstract}

 \maketitle

Most abundant among heavy non-volatile elements in the Universe,
iron has been extensively studied in its solid and liquid forms.
Where experiments are not feasible, computer modeling is used to
understand properties of liquid iron at high temperature and
pressure (see, e.g., Ref. \cite{alfe}) at which iron exists in the
core of planets such as Earth as well as giants such as Jupiter.
It is worth to note that even the location of melting line of iron
is not well established from experiment, but computer simulation
helps to clarify the experimental results \cite{premelting} Little
is known about iron properties in the supercritical region because
iron's critical point is located at severe (P,T) conditions. Yet
here, in the supercritical state, many interesting surprises
await, including the contemplated percolation-driven
metal-insulator transition (for review, see, e. g., Ref.
\cite{likalter}) and recently proposed dynamic, thermodynamic and
structural transitions
\cite{widominit,simeoni,phystoday,frenkel4,frenkel3,frenkel2,frenkel1,
frenkel5}.

Until recently, it has been held that no difference between a gas
and a liquid exists above the critical point. Instead, the
supercritical state has been viewed as qualitatively the same
everywhere on the phase diagram, with no distinct changes of
properties with varying pressure and temperature \cite{sbook}. Yet
the idea that the boiling line can be meaningfully extended above
the critical point has been long intriguing. For example, the
Widom line, the line of maxima of correlation length
\cite{widominit}, continues in the supercritical region and
vanishes not far above the critical point. In this approach, a
short bunch of lines, each corresponding to lines of maxima of
different physical properties such as heat capacity,
compressibility, thermal expansion and so on, starts from the
critical point and ends quite close to it \cite{nowidom,nowidom1}.
The alternative demarcation of the supercritical state, the
Frenkel line, is not related to the existence of the critical
point \cite{frenkel4}, and separates two physically distinct
states at arbitrarily high pressure and temperature above the
critical point
\cite{phystoday,frenkel4,frenkel3,frenkel2,frenkel1,frenkel5}.

The main idea behind the Frenkel line (FL) is that regardless of
how far one is above the critical point, the crucial difference in
atomic dynamics can still be ascertained. Accordingly, the FL
demarcates the transition between the liquid-like regime where
particle dynamics consists of both oscillations and ballistic
jumps between quasi-equilibrium positions and the gas-like regime
where the oscillatory component is lost and the motion is purely
ballistic as in the gas. This change in dynamics leads to a
dramatic change of all major system properties, including
diffusion, viscosity, thermal conductivity, speed of sound,
specific heat, structure as well as the ability to maintain
solid-like shear modes at any frequency
\cite{phystoday,frenkel4,frenkel3,frenkel2,frenkel1}. Since the
existence of shear modes is associated with rigidity, the
transition at the FL is the transition between the ``rigid''
liquid and ``non-rigid'' gas-like fluid.

Microscopically, the change of dynamics at the FL can be
unambiguously quantified by the disappearance of the minima in the
velocity autocorrelation function, signalling the loss of
oscillatory component of motion \cite{frenkel4}.

The FL has been calculated for model systems such as Lennard-Jones
(LJ) and soft-sphere liquids
\cite{frenkel4,frenkel3,frenkel3,frenkel2,frenkel1,frenkel5}. An
important question is whether one can propose a general and simple
recipe for drawing the FL on the phase diagram for any liquid,
without detailed system-specific calculations. In this paper, we
answer this question positively, and use iron, with its
non-trivial many-body interactions, as an important case study. We
subsequently discuss the relationship between the FL and
percolation transition in supercritical fluids as well as
transitions between conducting and insulating states. We finally
point the relevance of our results to the evolution of iron
conductivity and emergence of magnetic fields in planets and
exoplanets at various stages of their evolution.

We have used classical molecular dynamics (MD) simulations with
3456 atoms interacting via many-body embedded atom potential
\cite{belonoshko} optimized for the liquid iron. Liquids were
first equilibrated for 1 ns in the constant volume and temperature
ensemble, and subsequently used for production runs in the
constant energy ensemble with a smaller-than-usual timestep (0.2
fs) in order to properly resolve the velocity autocorrelation
function at high temperature. We have simulated several density
and temperature points in significant excess of the critical
point.

A representative example of the velocity-autocorrelation function
(VAF) is shown in Fig. 1a. We observe the transition between two
regimes with and without minima of the VAF, corresponding to the
loss of the oscillatory component of atomic motion
\cite{frenkel4}. This transition defines a point at the FL in
($P$,$T$) or ($\rho$,$T$) coordinates, where $P$ is pressure and
$\rho$ is density.

\begin{figure}
\includegraphics[width=7cm, height=7cm]{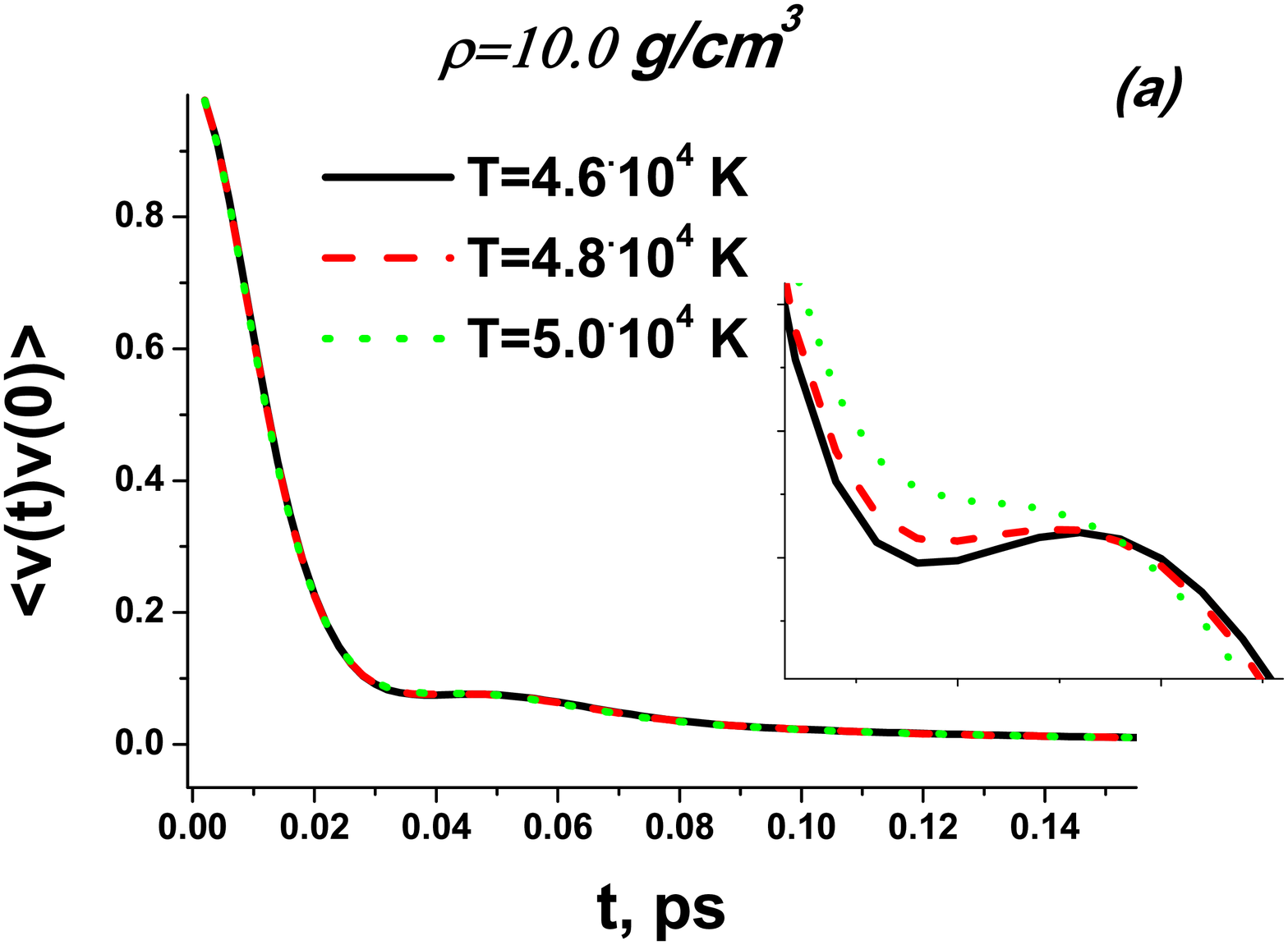}%

\includegraphics[width=7cm, height=7cm]{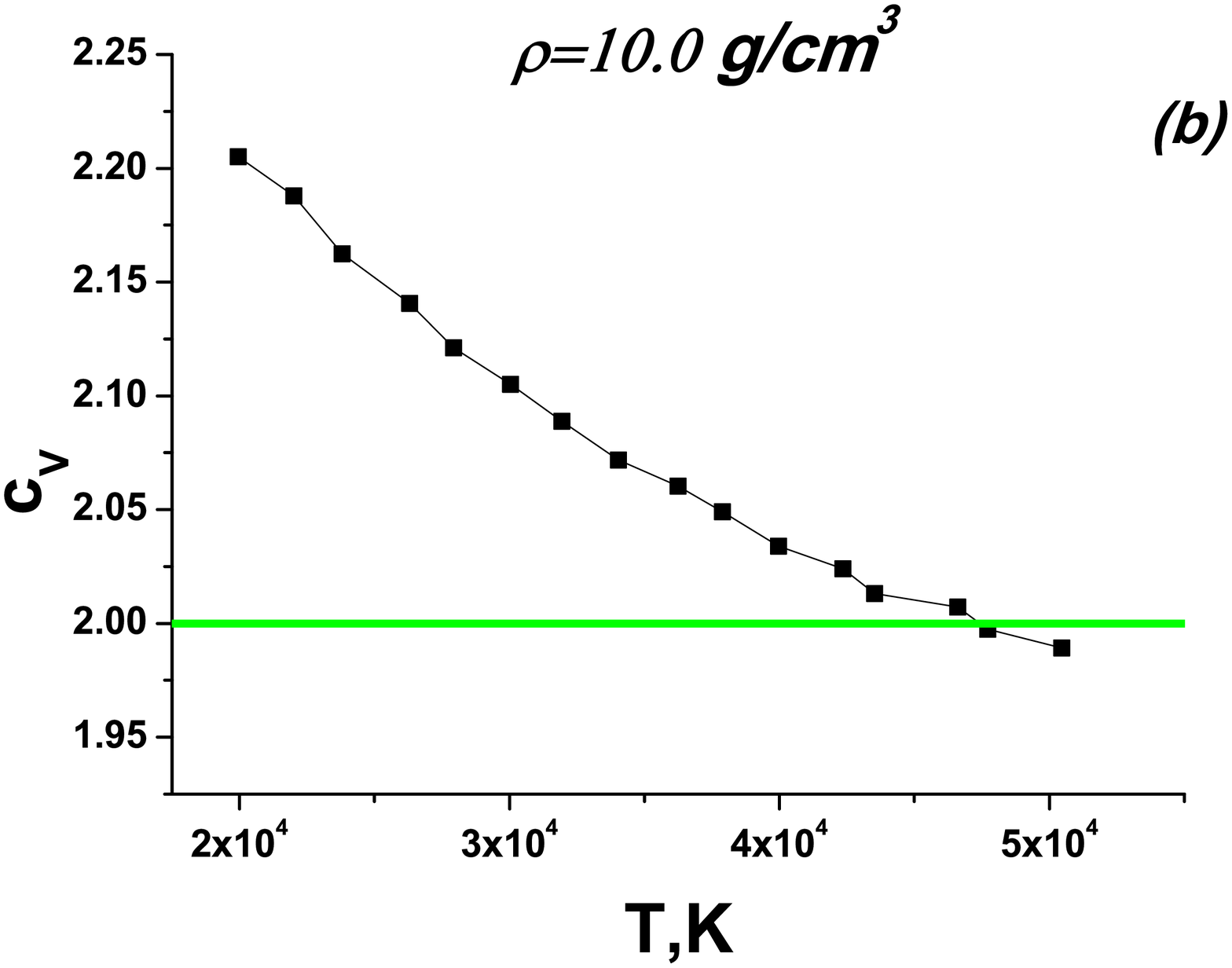}%

\caption{\label{fig:fig1} (a) Velocity autocorrelation functions
for iron at $\rho=10$g/cm$^3$ and different temperatures. The
inset enlarges the oscillating part of the plot. (b) Heat capacity
of iron at the same density (Color online).}
\end{figure}

In addition to the dynamic VAF criterion, the FL can be located
using the thermodynamic criterion: $c_v=2$ ($k_{\rm B}=1$).
Indeed, liquids support solid-like shear modes at frequency
$\omega>\frac{1}{\tau}$, where $\tau$ is liquid relaxation time,
the average time between atomic jumps at one point in space
\cite{frenkelhimself}. As discussed above, the FL corresponds to
the loss of oscillatory component of motion and
$\tau\approx\tau_{\rm D}$ ($\tau_{\rm D}$ is the shortest
vibration period), implying that the system loses its ability to
support shear modes at all available frequencies. This gives the
system energy of $2NT$ ($\frac{3}{2}NT$ kinetic energy and
$\frac{1}{2}NT$ potential energy of the remaining longitudinal
mode) and $c_v=2$ \cite{kt2008}. It is important to note that here
we neglect the anharmonic effects. Although it is not very
precise, it does not alter the qualitative behavior of the system.
In Fig. 1b, we show a representative $c_v$, calculated as
temperature derivative of the energy along an isochore.

In Fig. 2, we show the FL calculated using both dynamic and
thermodynamic criteria and find a very good agreement between the
two calculations. This serves as an important self-consistency
check. The agreement is similar to that we observed earlier for
the simple LJ system \cite{frenkel4}.

\begin{figure}
\includegraphics[width=7cm, height=7cm]{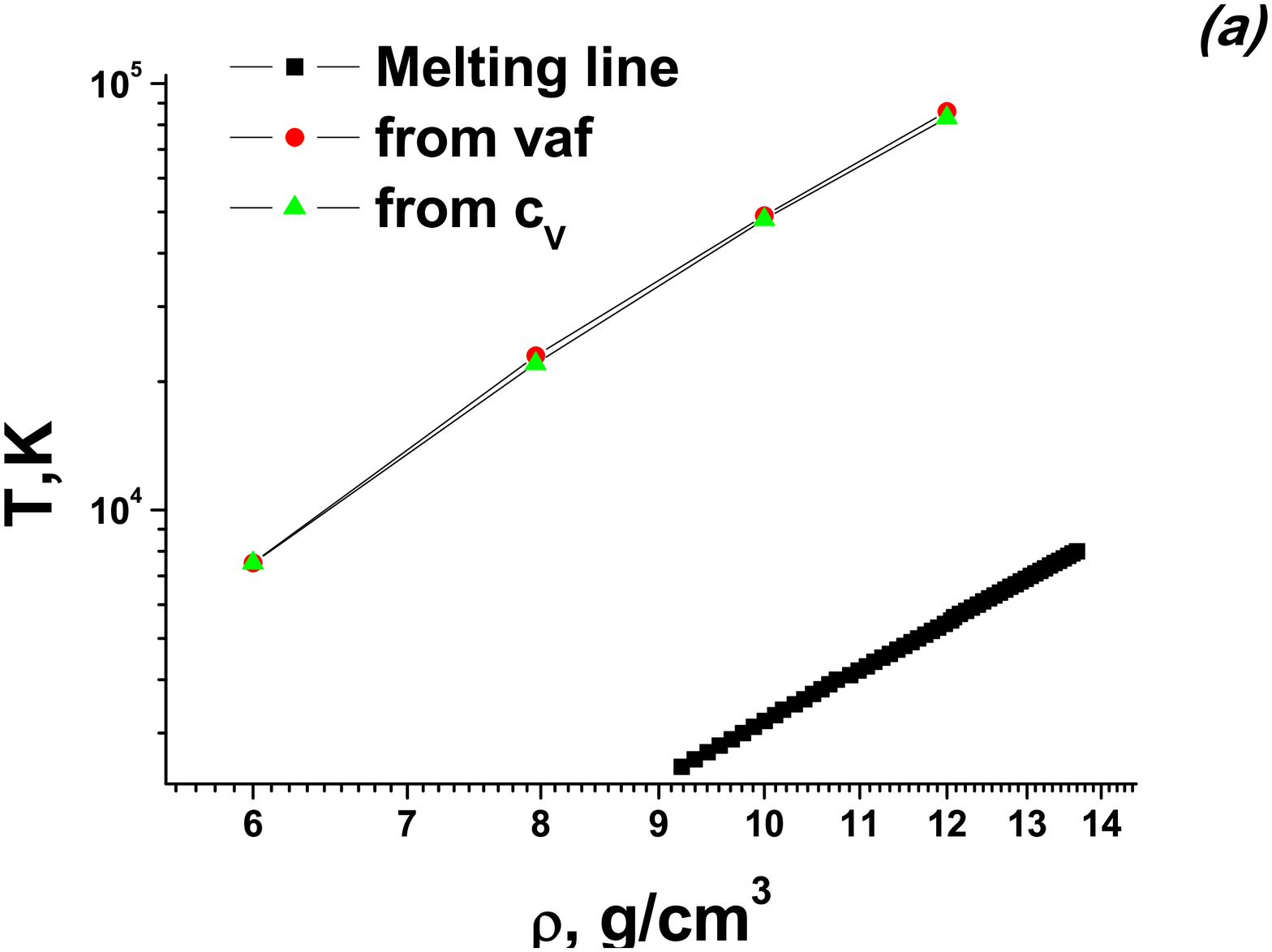}%

\includegraphics[width=7cm, height=7cm]{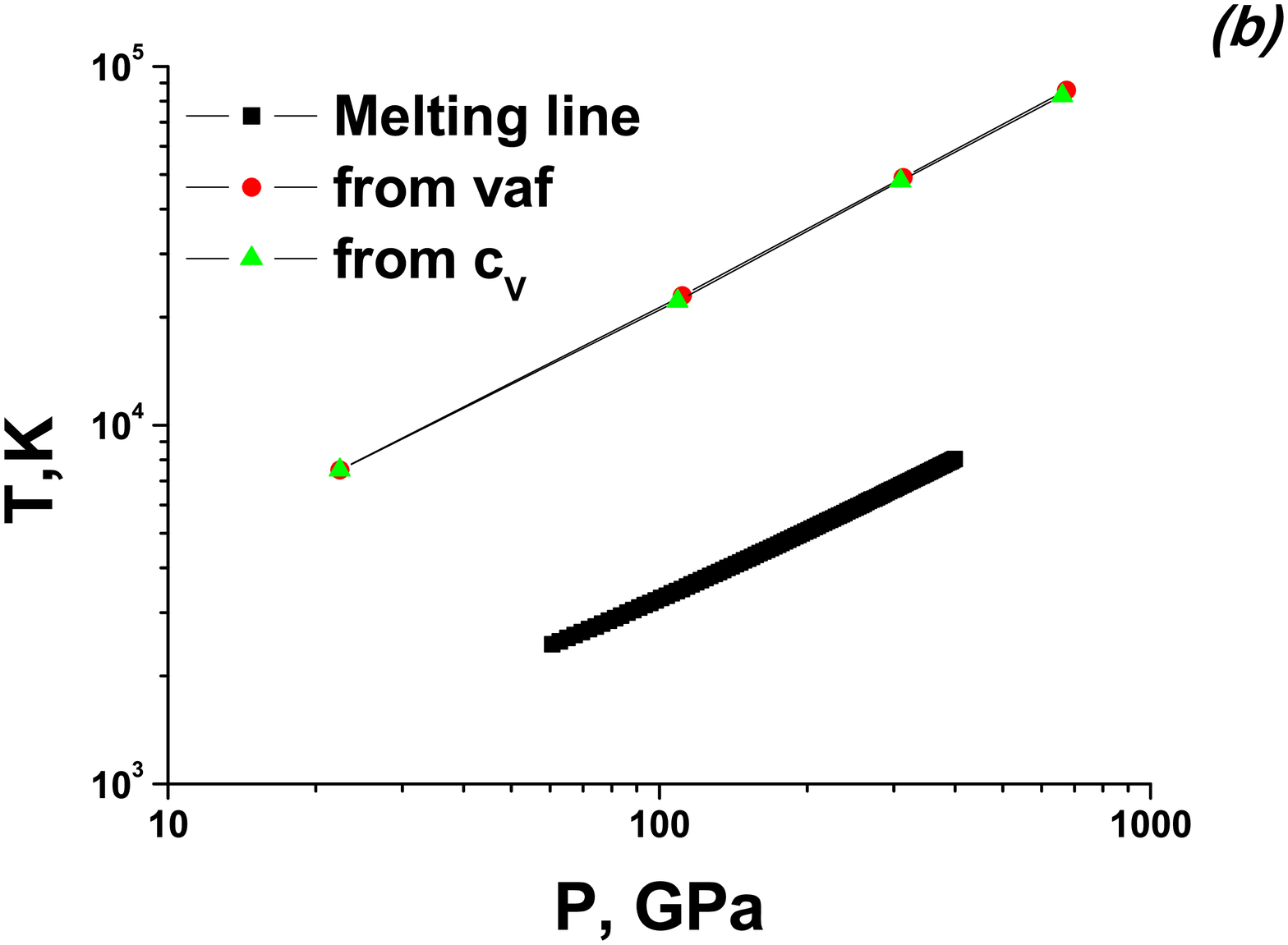}%

\caption{\label{fig:fig2} The Frenkel line of iron from two
criteria (see the text) in (a) $\rho-T$ and (b) $P-T$ coordinates.
(Color online).}
\end{figure}

In Fig. 3, we show the FL on the phase diagram together with the
melting line and the critical point. The melting line is plotted
using the data from previous calculations \cite{belonoshko}, and
the critical point is taken as $T_c=9250$ K (with uncertainty of
$1000$ K) and $\rho_c=2.0$ g/cm$^3$ \cite{crit,crit1}. The boiling
line has been calculated from the law of corresponding states and
previous data for the LJ system \cite{lj-lg}. We observe that the
FL crosses the boiling line at approximately $\rho \approx 4.5$
g/cm$^3$ and $T \approx 6500$ K, or $0.7T_c$ and $2\rho_c$, which
in units relative to the critical point is very similar to what
was previously found in the LJ system \cite{frenkel4}.

\begin{figure}
\includegraphics[width=7cm, height=7cm]{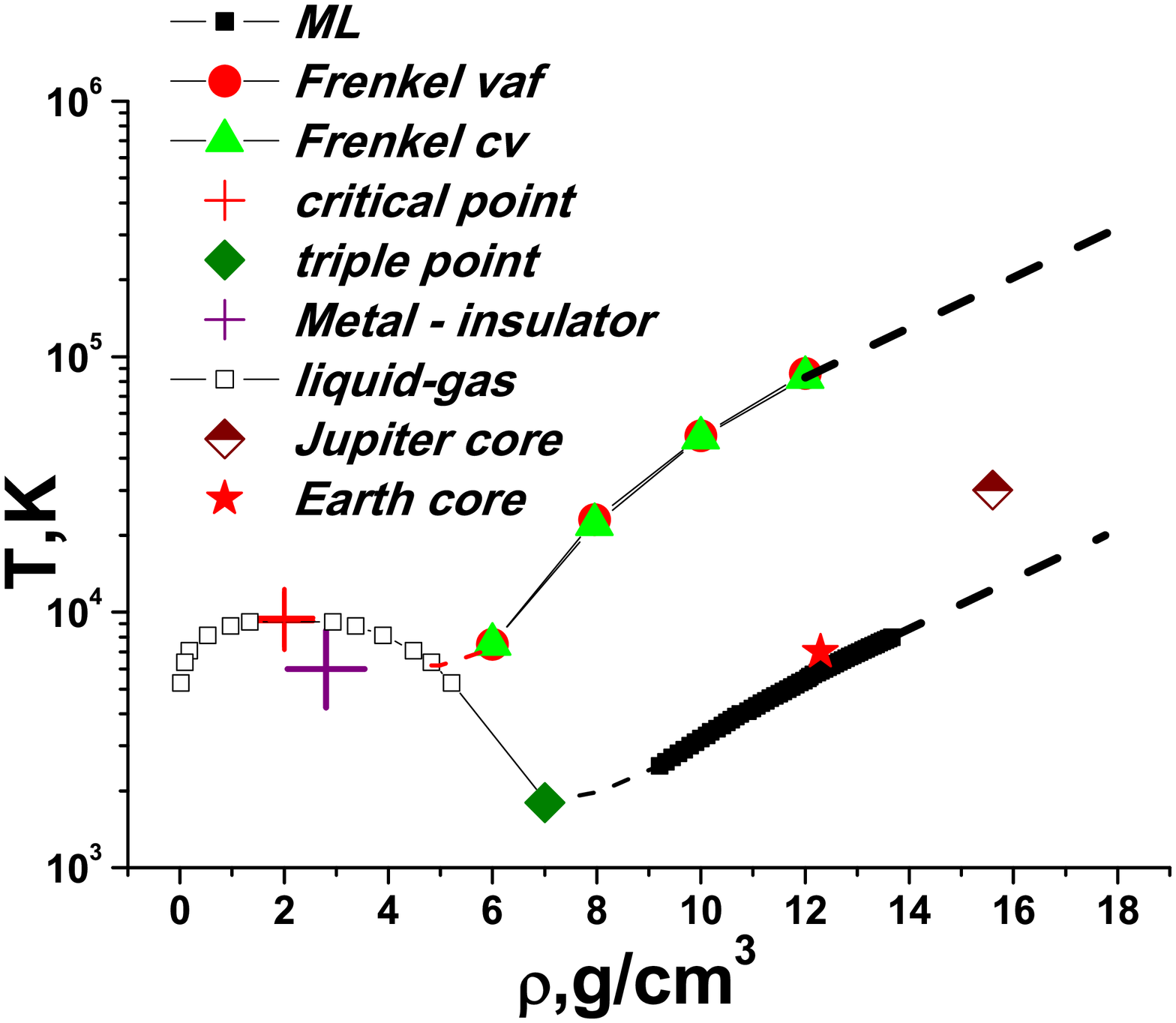}%

\caption{\label{fig:fig3} Placing Frenkel line on the phase
diagram of iron. Errors in determination of critical point and
metal to insulator transition point are of the order of the symbol
size. (Color online).}
\end{figure}

Locating the FL for iron is our first result. It immediately
enables one to predict the state of the supercritical liquid in
several cases of interest. For example, pressure and temperature
of iron in the Jupiter core is $P=90$ Mbar and $T=3.0\cdot 10^4$ K
\cite{jupiter}. Using our MD simulations, we have calculated the
corresponding density $\rho=15.6$ g/cm$^3$. We observe that the
location of this point on the phase diagram is below the FL (see
Fig. 3), implying that liquid iron in the center of Jupiter core
is in the rigid liquid state rather than in the gas-like non-rigid
fluid state, with all physical implications for the major
properties discussed above. The same is the case for iron in the
Earth core which lies even further down from the FL as compared to
Jupiter in Fig. 3. This is an important assertion for constructing
models of planet interiors and understanding essential effects
such as, for example, diffusion, viscosity, convection and thermal
conductivity \cite{jupiter}. Notably, this includes conductivity
and generation of magnetic field.

An interesting observation is that the rigid-liquid state of iron
in planets corresponds to current conditions only. At the early
stages of planet evolution, core temperatures were significantly
higher, implying that supercritical iron existed in the gas-like
non-rigid fluid state (see Fig. 3) where all major properties are
qualitatively different as discussed above. The same applies to
younger and hotter exoplanets. We therefore propose that as
planets and exoplanets cool off, supercritical iron in the core
may undergo the transition at the FL. One important consequence of
this transition is related to conductivity and magnetic fields.

We now explore whether the FL can be drawn for any liquid without
detailed system-specific simulations such as VAF or $c_v$. There
is a theoretical argument that after a certain pressure threshold,
the FL should be parallel to the melting line in the
double-logarithmic plot. Indeed, at high pressure the
intermolecular interaction is reduced to its repulsive part only
whereas the cohesive attracting part no longer affects
interactions. For the repulsive potential %$U\propto\frac{1}{r^n}$,
$U(r)= \varepsilon \left ( \frac{\sigma}{r} \right ) ^{n}$ (so
called soft spheres system) scaling of pressure and temperature is
well-known: system properties depend only on the combination of
$TP^\gamma$, where $\gamma$ is uniquely related to $n$.
Consequently, $TP^\gamma=$const on all ($P$,$T$) lines where the
dynamics of particles changes qualitatively. The qualitative
change of dynamics takes place on both FL and melting lines.
Therefore, the FL and melting lines are parallel to each other in
the double-logarithmic plot. This has been ascertained in direct
simulations for the simple LJ liquid \cite{frenkel4,frenkel3}.

The important question is whether the parallelism of the FL and
melting lines exists for systems where interactions are not
reduced to pair power-law interactions. Iron is an important
system in this respect: the embedded-atom potential we use
\cite{belonoshko} includes many-body terms so that the energy of
interaction between an atomic pair depends on the location of many
other atoms. Importantly, we observe in Figure 2 that the FL and
the melting lines are parallel in the double-logarithmic plot.
This implies that as far as scaling is concerned, interactions can
be well approximated by power-law dependence in a fairly large
pressure and temperature range despite the apparent complexity of
interactions as in iron \cite{weiron}.

Combined with the earlier results for the LJ and other simple
systems \cite{frenkel4}, this result enables us to propose a
simple and general way to locate the FL: it starts slightly below
the critical point (slightly below as mentioned above) and runs
parallel to the melting line at high pressure limit. We note at
this point that by its nature, the FL is not related to the
critical point and its overall existence. Indeed, we have recently
shown that the transition at the Frenkel line takes place in the
soft-sphere system where no boiling line and critical point exist
altogether \cite{frenkel4}.

We now focus on the relationship between the FL and atomic
arrangements in iron. Since early proposals \cite{ashcroft}, the
structure of liquid metals is considered to be well described by a
collection of hard spheres of some effective diameter
$d_{\textrm{eff}}$ at a given density. Several methods to
determine $d_{\textrm{eff}}$ were proposed that employ the
position of the first peak of structure factor $S(k)$. We find the
location of the $S(k)$ first peak along both the FL and the
melting line and show the corresponding $d_{\textrm{eff}}$ in Fig.
4. We observe that $d_{\textrm{eff}}$ decreases with temperature.
Interestingly, the packing fraction of spheres, $\eta =
\frac{\pi}{6} \rho d_{ \textrm{eff}}^3$, is constant along both
lines (see Fig. 5). We observe that at the melting line, $\eta
\approx 0.5$. This is close to the earlier result that melting in
the system of hard spheres corresponds to $\eta=0.49$
\cite{hsmelting}. This suggests a novel melting criterion: melting
corresponds to the packing fraction of hard spheres at the
hard-spheres melting line, an insights that warrants further
consideration.

\begin{figure}
\includegraphics[width=7cm, height=7cm]{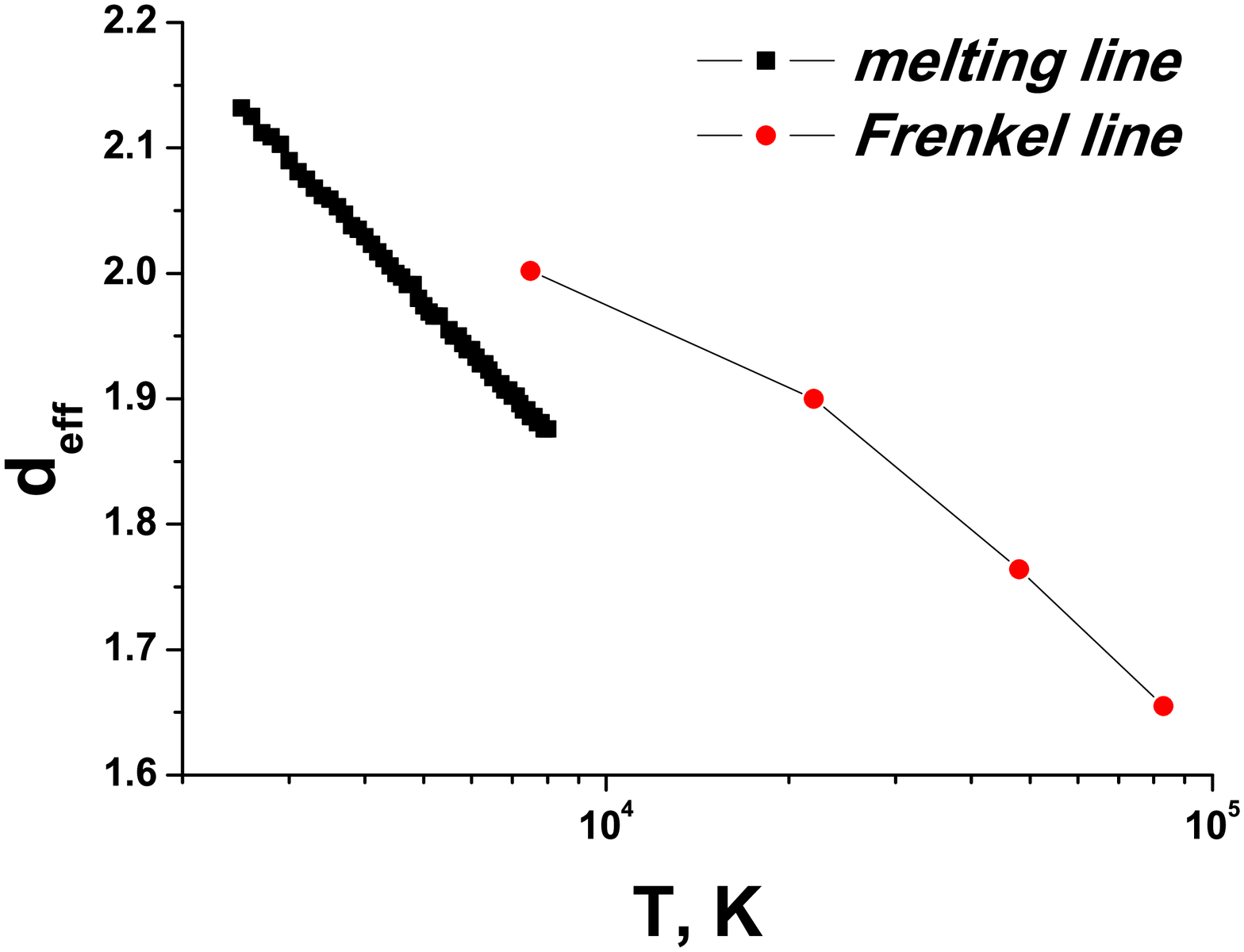}%

\caption{\label{fig:fig4} Diameter of effective hard spheres along
the melting line and the Frenkel lines of iron. (Color online).}
\end{figure}

\begin{figure}
\includegraphics[width=7cm, height=7cm]{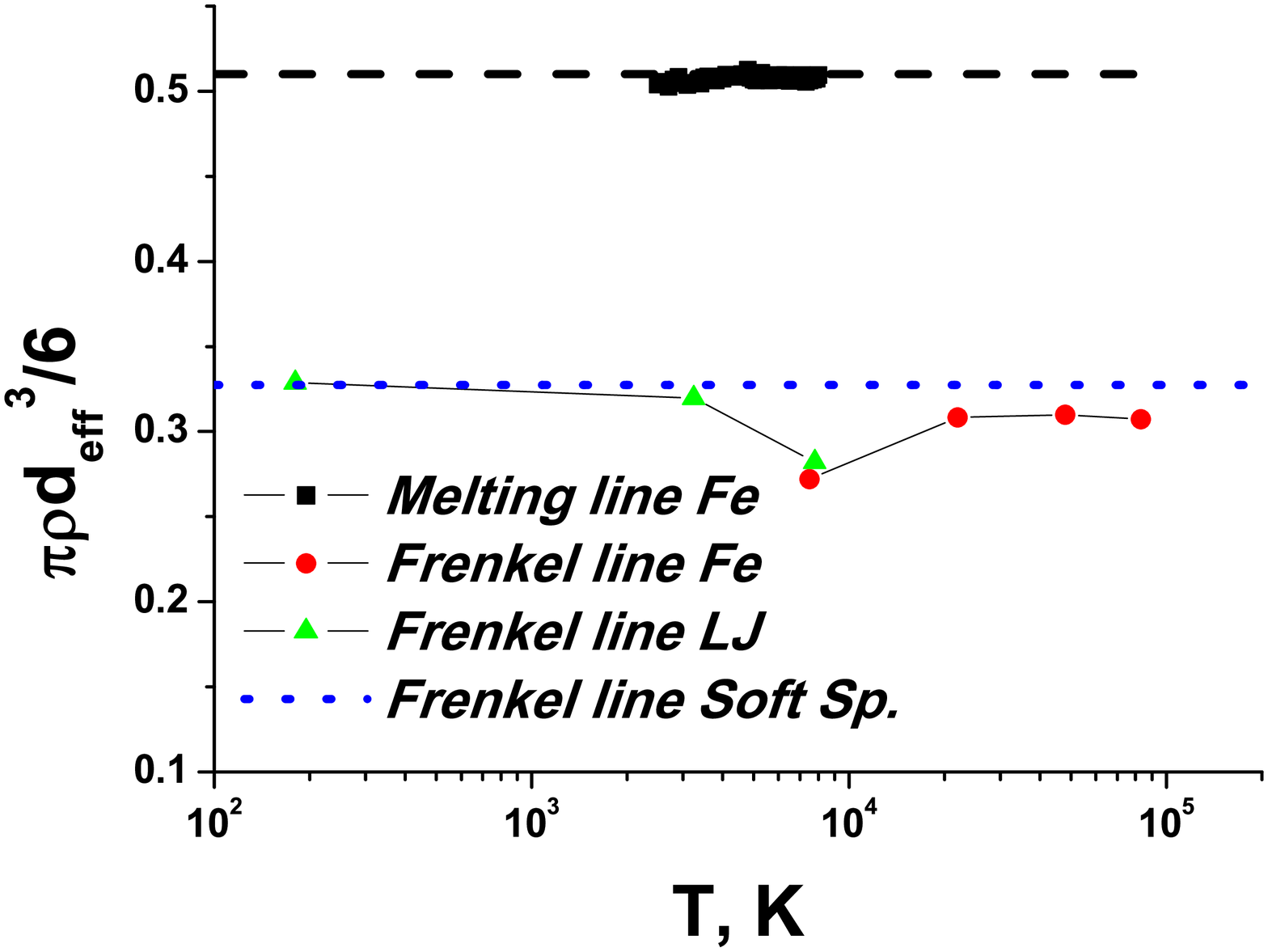}%

\caption{\label{fig:fig5} Packing fraction of effective hard
spheres along the melting and Frenkel lines of iron, along Frenkel
line of LJ fluid and along Frenkel line of soft spheres with
$n=12$. (Color online).}
\end{figure}

Importantly, we observe that $\eta \approx 0.3$ at the FL (see
Fig. 5). We have used our earlier data for the LJ liquid and soft
spheres with $n=12$ \cite{frenkel4}, and have repeated the same
procedure for calculating $\eta$ along the Frenkel line for these
systems. In Fig. 5 we observe that similar to iron, $\eta \approx
0.3$ for the LJ liquid at the FL.

%In our previous works we also reported Frenkel line for soft
%spheres system which is defined by interatomic potential $U(r)=
%\varepsilon \left ( \frac{\sigma}{r} \right ) ^{n}$. The most
%studied soft spheres system is the one with $n=12$.

In case of soft spheres phase diagram becomes effectively one
dimensional. Both melting and Frenkel lines are defined via
equations $\gamma =const$ with $\gamma = \rho \sigma ^3 \left(
\frac{\varepsilon}{k_BT} \right )^{3/n}$. Melting of soft spheres
with $n=12$ corresponds to $\gamma_m=1.15$ \cite{hoover}, while
Frenkel line is characterized by $\gamma_F=0.577$. Barker diameter
of soft spheres can be expressed as

\begin{equation}
  d_{B,soft}= \sigma \left ( \frac{\varepsilon}{k_BT}
  \right)^{1/n}\Gamma(\frac{n-1}{n}).
\end{equation}

From this formula one can see that packing fraction of effective
hard spheres $\eta= \frac{\pi}{6} \rho d_{B,soft}^3$ is simply
proportional to $\gamma$ and constant along the Frenkel line. The
magnitude of $\eta$ at the Frenkel line is $\eta_{F,n=12}=0.327$
which is also very close to the results for LJ and iron.

The importance of $\eta \approx 0.3$ is that it corresponds to the
appearance of the percolating cluster in the system of overlapping
spheres \cite{likalter}. This coincidence is meaningful in the
following sense. Below the FL, the presence of oscillatory
component means that atomic clusters can be well-defined, and
typically exist during liquid relaxation time $\tau$. For example,
the recent work identified the abundance of different stable
clusters in a high-temperature liquid, and further established
that the concentration of the tetrahedral clusters at the FL is
15--17 \%, close to the percolation threshold in the system
composed from tetrahedra \cite{kolyaroma}. In contrast, above the
FL where the oscillatory component of motion is lost and atoms
move ballistically as in a gas, the concentration of effective
spheres does not reach the percolation limit.

The latter circumstance is important for understanding
metal-insulator transitions in liquids. These transitions are
based on the percolation transition in the system of charged
effective hard spheres (for review, see, e.g., Ref.
\cite{likalter}). Here, metallic and insulating states in the
liquid emerge at packing fractions above and below the percolating
threshold, respectively. It is therefore important that the
experimental metal-insulator transition in liquid iron
($\rho=2.8\pm0.5$ g/cm$^3$, $T=6500\pm1500$ K
\cite{metins1,metins4}) is located not far from the FL (see Fig.
3). We note that the metal-insulator transition has also been
observed in liquid mercury \cite{likalter}. Once the FL line in
mercury is drawn similarly to that in Fig. 3 and following our
general recipe for locating the FL discussed above, the proximity
of the FL to the experimental metal-insulator transition becomes
apparent \cite{likalter}. One should note, however, that due to
high degree of ionization nonmetallic state of liquid iron still
has quite high conductivity \cite{metins1,metins4}.

We can now assert the state of conductivity of supercritical iron
at different pressure and temperature conditions. Using the above
density of iron in the core of Jupiter, we calculate $\eta=0.4$,
and find that $\eta$ is above the percolating threshold of 0.3. We
therefore predict that iron in the Jupiter core is conducting.

We similarly find that $\eta=0.5$ in the Earth core corresponds to
the highly conducting core, the result consistent with current
understanding. However, we noted above that in the early hotter
Earth, iron in the core could be above the FL, and therefore
existed in the low conducting state. This has important
consequence for understanding of the evolution of the Earth's
magnetic field: the magnetic field emerged at early stages of
Earth evolution when cooling.

In summary, we proposed a general recipe to locate the Frenkel
line in the supercritical region of the phase diagram using iron
as an important case study. We discussed the relationship between
the FL and the metal-insulator transition in supercritical liquid
metals. This enables predicting the conducting state of iron in
Jupiter and smaller planets such as Earth. We further discussed
the evolution of core conductivity and associated evolution of
magnetic fields: as planets and exoplanets cool off, supercritical
core undergoes the transition to the rigid-liquid conducting state
at the Frenkel line.

\begin{acknowledgments}
Y. F. thanks the Russian Scientific Center at Kurchatov Institute
and Joint Supercomputing Center of Russian Academy of Science for
computational facilities. The work was supported in part by the
Russian Foundation for Basic Research (Grants No 14-02-00451,
13-02-12008, 13-02-00579, and 13-02-00913), the Ministry of
Education and Science of Russian Federation (project
MK-2099.2013.2) and  grant of the Government of the Russian
Federation 14.A12.31.0003.

\end{acknowledgments}


\end{document}